\documentclass[conference]{IEEEtran}
\pdfoutput=1

\usepackage{slashbox}
\usepackage[utf8]{inputenc}
\usepackage{cite}
\usepackage{graphicx}
\usepackage{url}
\usepackage{color}
\usepackage[nounderscore]{syntax} 
\usepackage{listings}
\usepackage{xspace}

\newcommand{\prodiden}[1]{\textit{$\langle$#1$\rangle$}\xspace}
\newcommand{\sd}{spreadsheet\xspace}
\newcommand{\sds}{spreadsheets\xspace}
\newcommand{\matlab}{\textsc{Matlab}\xspace}

\begin{document}

\title{Towards the Design and Implementation of Aspect-Oriented
  Programming for Spreadsheets}

\IEEEoverridecommandlockouts

\author{\IEEEauthorblockN{Pedro Maia\IEEEauthorrefmark{1}\IEEEauthorrefmark{2},
Jorge Mendes\IEEEauthorrefmark{1}\IEEEauthorrefmark{2},
Jácome Cunha\IEEEauthorrefmark{1}\IEEEauthorrefmark{3},
Henrique Rebêlo\IEEEauthorrefmark{4},
João Saraiva\IEEEauthorrefmark{1}\IEEEauthorrefmark{2}}
\IEEEauthorblockA{\IEEEauthorrefmark{1}HASLab / INESC TEC, Portugal}
\IEEEauthorblockA{\IEEEauthorrefmark{2}Universidade do Minho, Portugal\\
\url{{pedromaia,jorgemendes,jas}@di.uminho.pt}}
\IEEEauthorblockA{\IEEEauthorrefmark{3}Universidade Nova de Lisboa, Portugal\\
\url{jacome@fct.unl.pt}}
\IEEEauthorblockA{\IEEEauthorrefmark{4}Universidade Federal de Pernambuco, Brazil\\
\url{hemr@cin.ufpe.br}}
}

%
%

\maketitle

%
%

\begin{abstract}

A spreadsheet usually starts as a simple and single-user software artifact, but,
as frequent as in other software systems, quickly evolves into a complex system
developed by many actors.  Often, different users work on different aspects of
the same spreadsheet: while a secretary may be only involved in adding plain
data to the spreadsheet, an accountant may define new business rules, while an
engineer may need to adapt the spreadsheet content so it can be used by other
software systems. Unfortunately, spreadsheet systems do not offer modular
mechanisms, and as a consequence, some of the previous tasks may be defined by
adding intrusive ``code'' to the spreadsheet.

In this paper we go through the design and implementation of an aspect-oriented
language for spreadsheets so that users can work on different aspects of a
spreadsheet in a modular way. For example, aspects can be defined in order to
introduce new business rules to an existing spreadsheet, or to manipulate the
spreadsheet data to be ported to another system.  Aspects are defined as
aspect-oriented program specifications that are dynamically woven into the
underlying spreadsheet by an aspect weaver. In this aspect-oriented style of
spreadsheet development, different users develop, or reuse, aspects without
adding intrusive code to the original spreadsheet. Such code is added/executed
by the spreadsheet weaving mechanism proposed in this paper.
\end{abstract}

%
\section{Introduction}
\label{sec:introduction}

Spreadsheet systems are the software system of choice for many
non-professional programmers, often called end-user programmers~\cite{enduser}, like
for example, teachers, accountants, secretaries, engineers, managers,
etc. In the last century such end users would develop their
spreadsheets individually in their own desktops, and sharing and reuse
was not the usual procedure.  The recent advent of powerful mobile
devices, and, as a consequence, the availability of powerful cloud-based
spreadsheet systems (like, for example, Google Drive), has
dramatically changed this situation. Nowadays, spreadsheets are
complex software systems, developed and maintained by many (end) users.

Very much like in the development of other software systems, different
developers (end users in this case) are concerned with
different aspects of the functionality of the
system. However, while modern programming languages offer modularity
mechanisms providing powerful abstractions to develop software
collaboratively, spreadsheet systems offer no such support to their
users. As a result, a spreadsheet tends to evolve into a single
software artifact where all business logic from all different users is
defined! In such a collaborative setting if a new user needs to
express a new business rule on the spreadsheet data, he/she has to do it
by intrusively adding formulas to the existing spreadsheet.

The programming language community developed advanced modularity
mechanisms to avoid this problem in regular programming languages.
In that sense, aspect-oriented programming (AOP) is a popular and advanced
technique that enables the modular implementation of the so-called 
crosscutting concerns. The crosscutting structure tends to appear
tangled and scatted across several artifacts of a software system. 
While implementing such crosscutting concerns, the crosscutting 
structure can appear in common software development concerns,
such as distribution and persistence~\cite{Soares-Laureano-Borba02}, 
error handling~\cite{Lippert-Lopes00,Castor-etal09}, certain 
design patterns~\cite{Hannemann-Kiczales02}, tracing~\cite{Kiczales-etal01}, 
or design by contract~\cite{Kiczales-etal01,Rebelo-etal14,Rebelo-Lima-Leavens11}.

In this paper we introduce the concept of AOP to \sds.
We start by introducing a running example in Section~\ref{sec:motivation}.
Our first contribution is presented in Section~\ref{sec:architecture}
where we adapt the AOP concept to spreadsheets. For instance,
a key feature of a common aspect-oriented language is
the possibility to add advice before, after or around a join point.
Since \sds are two-dimensional, AOP features like advising need to 
be adapted in order to be applied to \sds.
The second contribution we do in this work is the design of a new
language to implement AOP for spreadsheets.
This language is presented in Section~\ref{sec:language}.
An overview of the architecture of the system is presented
in Section~\ref{sec:weaver}. Finally, Section~\ref{sec:related} discusses related work,
and Section~\ref{sec:conclusion} presents our conclusions.

%
\section{Motivation}
\label{sec:motivation}

\begin{figure}[ht]
	\centering
  \includegraphics[scale=0.5]{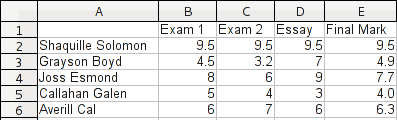}
  \caption{Student grading spreadsheet.}
  \label{fig:example}
\end{figure}

In this section we present an example of a popular setting to use spreadsheets:
manage the students marks in a course. Such a spreadsheet is shown in
Fig.~\ref{fig:example}.

In this simple spreadsheet, the final mark of a student is the average
of three evaluation items: two exams and one essay. Thus, this mark is
obtained by the formula \texttt{=AVERAGE(B2:D2)} (for the student in
line~2).

Let us consider now a real scenario where three different users
share, access, and update this spreadsheet: a \textit{teaching
assistant}, that mainly structures the spreadsheet and compiles the
different marks, the \textit{teaching coordinator} who has to validate
the spreadsheet, decide on borderline cases, and send the final
marks to the academic services, and finally, the \textit{Erasmus
coordinator} who has to adapt the marks of Erasmus students to the
ECTS system\footnote{For more details on the ECTS system please refer to
\url{http://ec.europa.eu/education/tools/docs/ects-guide_en.pdf}.}.

Because spreadsheets offer no modularity mechanisms, a typical spreadsheet
development environment allows a spreadsheet to quickly evolve into a more
complex one. That is, each of the users adds/updates data and
formulas in order to express the (crosscutting) logic they need. In the following, 
we show the spreadsheet after the teaching and Erasmus coordinator update it.

\begin{figure}[ht]
	\centering
  \includegraphics[scale=0.5]{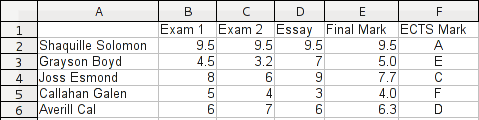}
	\caption{Example of a student grading spreadsheet with mark adjustments (cell
	E3) and ECTS marks (column F).}
  \label{fig:example-round-ects}
\end{figure}

The data edited by one user can be seen as intrusive code for the
others. For example, the teaching coordinator and its assistant are
not concerned with the ECTS system, although this concern is
now part of their spreadsheet.

Let us analyze the teaching coordinator role in more detail. Because
he decides on borderline cases, he decided to approve only one of such
cases (student in line~3). Instead of changing that particular final
mark formula, and as often occurs in reality, he just replaces this
formula by the constant \texttt{5.0}. Although the result of such an
operation is a correct spreadsheet, this action has several problems.
First, there is no documentation on how borderline cases have been
decided (the threshold is not known). Second, the original spreadsheet
data is lost, and it may be difficult to recover it (for
example, if other borderline students ask for explanations).

This is the natural setting to apply AOP. In this style of programming
users do not intrusively modify the original program (a spreadsheet in
our case), but instead, a new software fragment is defined in order to
specify the program transformations needed to express such new
concern. These software fragments are called \textit{aspects}. Then, a specific
AOP mechanism, called \textit{weaver}, given the original program and the
aspect(s), weaves them into a coherent executable software program.

We present in Listing~\ref{lst:borderline} our first spreadsheet aspect, the one that
specifies the coordinator concerns on borderline cases.

\begin{lstlisting}[caption={An aspect to handle borderline marks.},label={lst:borderline}]
aspect BorderlineCase
finalmark : select sheet{*}.column{*}.cell{*}
around finalmark {
    #{cell.result >= 4.8 && cell.result < 5 ? 5 : cell.value}
  } when {
    cell.column[0].value = "Final Mark"
  }
end
\end{lstlisting}

An aspect-oriented language has three main parts. First it is necessary to select
the \textit{join points} of interest by means of pointcut declarations.
In the example, the pointcut \texttt{finalmark} selects any cell within
any worksheet. This is done by the \textit{select} command in our language.
Then, the \textit{around} advice declaration defines the actions
(transformations in our case) to be applied. 
Hence, it specifies when the \textit{result} of evaluating a formula is greater
than of equal to \texttt{4.8}, and if it is, the cell is replaced by the constant \texttt{5.0}.
To access the result of the computation of a cell, that is, the result
of the formula in the cell, we use \lstinline|cell.result|. This accesses
a cell, and after that, its computed \textit{result}.
The \textit{when} statement specifies when the action is applied (in
this case to all columns labeled \texttt{Final Mark}).

This aspect clearly and non-intrusively defines the crosscutting rule used
to decide on borderline cases: all students with marks greater
than or equal to \texttt{4.8} are approved in the course.

Similarly, the Erasmus coordinator can define an aspect where the
rules to define the ECTS marks are expressed. 
This aspect is presented in Listing~\ref{lst:ectsmark}.

\begin{lstlisting}[caption={Aspect to add ECTS marks.},label={lst:ectsmark}]
aspect AddECTSMark
finalmarks : select sheet{*}.column{*}.cell{*}

right finalmarks {
    =IF(#{cell.name}<=10 && #{cell.name}>=9.5
      , "A"
			, IF(#{cell.name}<9.5 && #{cell.name}>=8.5
        , "B"
				, IF(#{cell.name}<8.5 && #{cell.name}>=6.5
          , "C"
					, IF(#{cell.name}<6.5 && #{cell.name}>=5.5
            , "D"
						, IF(#{cell.name}<5.5 && #{cell.name}>=5
              , "E"
              , "F")))))
	} when {
    column[0].value == "Final Mark"
      && cell.row <> 0
  }

right finalmarks {
    ECTS Mark
  } when {
    column[0].value = "Final Mark"
      && cell.row = 0
  }
end
\end{lstlisting}

In this aspect, we use \lstinline|cell.name| to access the cell
address (for instance ``E4''). Thus, when the \lstinline|IF| formula
is placed in the corresponding cells, the correct references
will be calculated and inserted in the formula replacing the 
use of \lstinline|cell.name|.

An aspect weaver can then weave this aspects into the original
spreadsheet individually, or by composing them. Actually,
Fig.~\ref{fig:example-round-ects} shows the result of weaving
the aspect \textit{AddECTSMark} after the
aspect \textit{BorderlineCase}. Thus, aspects can be combined. It
should also be noticed that such a spreadsheet based weaver has to
dynamically weave aspects since for some aspects
(\textit{BorderlineCase}) only after computing formulas, aspects can
be weaved. Indeed, it is necessary to execute the spreadsheet
to get the results of formulas, for instance, when the \lstinline|cell.result|
operator is used.

We have just presented two aspects of our AOP spreadsheet language and
briefly explained how aspects are weaved into a spreadsheet system. Next
sections present in detail the design and implementation of this
language and weaver.

\section{Applying Aspect-Oriented Programming\\to Spreadsheets}
\label{sec:architecture}

In this section, we discuss how AOP concepts can be applied to spreadsheets. 
Specific examples are described in Section~\ref{sec:language}.

%
%
\subsection{Spreadsheet Join Point Model}

In order to apply AOP to a language, it is necessary to identify the
\textit{join point} model that the new aspect language supports. A join point is
a well-defined point in the program that is specified by a \textit{pointcut},
that is, an expression to match specific elements within the language.  In our
case, these elements are set to be the main elements of the spreadsheet, where
users will want to perform operations on, for instance, use an alternate
worksheet for testing or add more cells for debugging. For spreadsheets, we
define the following join points of interest:
\begin{itemize}
	\item worksheets;
	\item ranges; and,
	\item cells.
\end{itemize}

With these join points the spreadsheet is accessible from aspects, hence users
can separate concerns at different levels:
\begin{center}
\includegraphics[width=0.35\columnwidth]{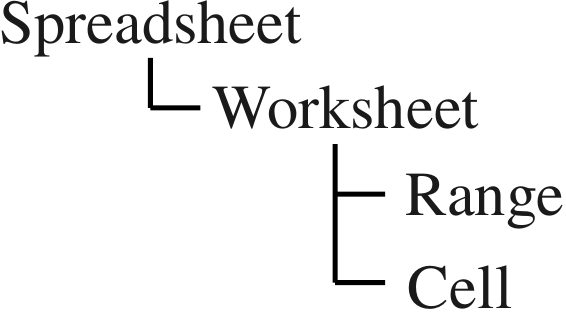}
\end{center}

When a join point is instantiated, it is possible to perform an action using
advice. Advice are additional behavior that one wants for the underlying program, that can
be new worksheets, ranges of cells, or single cells. This allows to separate the
business logic of the spreadsheet into several concerns when developing the
spreadsheet and then join everything in order to achieve the wanted application.

%
%
\subsection{Worksheet}

A spreadsheet file is composed of a set of worksheets, each of which containing
the cells with the data and formulas. Worksheets are the top-level artifacts in
a \sd.

\begin{figure}[ht]
  \centering
  \includegraphics[width=0.7\columnwidth]{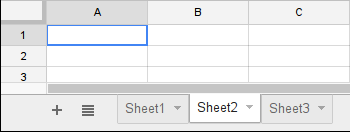}
  \caption{List of three worksheets within a spreadsheet file.}\label{fig:sheets}
\end{figure}

As a join point, a worksheet (see Fig.~\ref{fig:sheets}) can be modified by the
standard AOP advice in the following manner:
\begin{itemize}
\item {\bf before} -- insert a worksheet before the current join point;
\item {\bf after} -- insert a worksheet after the current join point;
\item {\bf around} -- insert a worksheet before and/or after the
                      current joint point and/or define an alternative
                      worksheet for the current join point.
\end{itemize}

As an example, consider Fig.~\ref{fig:sheets} with \texttt{Sheet2} as the join
point. A \textit{before} advice for this join point results in a new worksheet
between the worksheets \texttt{Sheet1} and \texttt{Sheet2}. On the other hand,
if we consider an \textit{after} advice, the resulting worksheet is between
worksheets \texttt{Sheet2} and \texttt{Sheet3}.

To specify a worksheet, two options are available: either defining the cells for
the new worksheet, or referencing a worksheet to be duplicated.

%
%
\subsection{Cell}

Cells are the finest grained join points possible in the spreadsheets world.
Unlike common programming languages, they are inserted in a two-dimensional
plan. Thus, the usual \textit{before} and \textit{after} advice declarations are
not completely adequate since ambiguities may arise about which of the two
dimensions (vertical or horizontal) should be used.  To overcome this issue,
each kind of advice declaration is separated in two for each dimension in the
plane, resulting in the following:
\begin{center}
\begin{tabular}{|l|l|l|}
\hline
\bf \backslashbox{direction}{advice} & \it \textbf{before} & \it \textbf{after} \\\hline
\textbf{vertical}   & above  & below \\\hline
\textbf{horizontal} & left   & right \\\hline
\end{tabular}
\end{center}

The advice that can be defined for cell join points is:
\begin{itemize}
  \item {\bf left} -- add a cell, or range of cells, to the left of the current
    join point;
  \item {\bf above} -- add a cell, or range of cells, above the current join
    point;
  \item {\bf right} -- add a cell, or range of cells, to the right of the
    current join point;
  \item {\bf below} -- add a cell, or range of cells, below the current join
    point;
  \item {\bf around} -- define an alternative cell for the current join point.
\end{itemize}

\begin{figure}[ht]
  \centering
  \includegraphics[width=0.7\columnwidth]{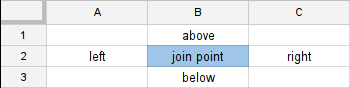}
  \caption{Two-dimensional advice.}\label{fig:advices2d}
\end{figure}

A cell is specified just by stating its contents. A range of cells is specified
by defining the cells that it contains.

%
%
\subsection{Range}

A range is a set of cells contained in a worksheet. As it regularly happens in
\sds, our setting only handles rectangular ranges, that is, the range must be a
rectangle and all the cells in that rectangle must part of the range.

Advice declaration for ranges is similar to the ones for cells, but special
care must be taken to match range sizes when adding new cells, or ranges before
or after it. Thus, we have the following possible advice declarations:
\begin{itemize}
  \item {\bf left} -- add a range of cells, to the left of the current join
    point;
  \item {\bf above} -- add a range of cells, above the current join point;
  \item {\bf right} -- add a range of cells, to the right of the current join
    point;
  \item {\bf below} -- add a range of cells, below the current join point;
  \item {\bf around} -- define an alternative range for the current join point.
\end{itemize}

%
\section{Language to Specify Aspects for Spreadsheets}
\label{sec:language}

In this section we present our language for aspects for spreadsheets.  This
language is based on existing ones for common purposes programming languages,
and implements the vision we described in Section~\ref{sec:architecture}.  The
proposed language allows to specify aspects defining its pointcuts and advice.
For each component of the aspects to be written by the user we present next the
corresponding grammar. This grammar is the artifact used to validate, though a
compiler, the correctness of the aspects written by the users.  We start by
introducing the grammar for pointcuts.

%
%
\subsection{Pointcuts}

Pointcuts are defined by an expression pattern, which is
specified by the following grammar:

\medskip
\begin{grammar}
<join\_point> ::= <jp\_name> `:' `select' <pExpr>
\end{grammar}
\medskip

\noindent
where \prodiden{pExpr} is a pattern expression. The expression is used to
define the kind of join point to be selected: \emph{sheet}, \emph{range}, or
\emph{cell}.

The allowed expressions in join points are instances of the
following productions:

\medskip
\begin{grammar}
<pExpr> ::= <pSheet>
\alt <pSheet> `.' <pRange>
\alt <pSheet> (`.' <pRange>)? `.' <pCell>

<pSheet> ::= `sheet' `{' <pSheetExpr> `}'

<pSheetExpr> ::= `name' <bComp> <string>
\alt `number' <bComp> <integer>
\alt `*'

<pRange> ::= `range' `{' <pRangeExpr> `}'
\alt `column' `{' <pRangeExpr> `}'
\alt `row' `{' <pRangeExpr> `}'

<pRangeExpr> ::= `name' <bComp> <string>
\alt `*'

<pCell> ::= `cell' `{' <pCellExpr> `}'

<pCellExpr> ::= `name' <bComp> <string>
\alt `match' <bComp> <string>
\alt `*'

<bComp> ::= `==' | `<' | `<=' | `>' | `>=' | `<>'
\end{grammar}
\medskip

The pattern expression \prodiden{pExpr} allows to select worksheets
\prodiden{sheet}, ranges \prodiden{range}, or cells \prodiden{cell}. 

When
specifying a worksheet as a join point, it is possible to select a worksheet relative
to a given name, to a worksheet index, or to select all worksheets. 

For ranges,
there are three kinds that can be selected: a column range (\texttt{\small
column}) with a width of one cell, a row range (\texttt{\small row}) with a
height of one cell, or a range with any rectangular shape (\texttt{\small range}). 

For cells, it is possible to specify its address (\texttt{\small name}), or a pattern-match expression
(\texttt{\small match}). When
no specific worksheet, range or cell is necessary, the wildcard symbol \lit{*} can be used.

With this, we can select, for instance, the second worksheet

\medskip
\lstinline|worksheet_jp: select worksheet{number=2}|
\medskip

\noindent
or a rectangular range of 3 columns by 2 rows starting at cell \texttt{A2} in any
worksheet:

\medskip
\lstinline|range_jp: select worksheet{*}.range{name="A2:C3"}|
\medskip

\noindent
or the cells in the first row of any worksheet:

\medskip
\lstinline|cell_jp: select worksheet{*}.range{row=1}.cell{*}|
\medskip

Depending on the kind of join point selected, different artifacts are made
available to work with within the advice. For worksheets, we can use the variable
\lstinline|worksheet|, and then one of the attributes: \texttt{\small name} or
\texttt{\small number}. For ranges, depending on its kind, the available
variable can be: \texttt{\small range}, \texttt{\small column}, or
\texttt{\small row}; they have the attribute \texttt{\small name}.
Moreover, since ranges are sets of cells, indexation can be used, for instance, to
select the first row of a column, one can write \texttt{\small column[0]}. For
cells, we have the variable \texttt{\small cell} which has the attribute
\texttt{\small name} (its cell reference).

%
%
\subsection{Advice}

Advice are the actions to apply to the join points. They are defined by the
following grammar:

\medskip
\begin{grammar}
<advice> ::= $\left(<advice\_name> `:'\right)?$ <advice\_position> <jp\_name> `{' <code> `}' <advice\_condition>

<advice\_position> ::= `left' |~`above' |~`right' |~`below' |~`around'

<advice\_condition> ::= $\left(`when {' \;  <bExpr> \; `}'\right)?$
\end{grammar}
\medskip

\noindent
where \prodiden{code} is a cell or a list of cells and respective contents:

\medskip
\begin{grammar}
<code> ::= <string> | <cellList>

<cellList> ::= <cellRef> `=' <string> $\left(`;' <cellList>\right)?$
\end{grammar}
\medskip

The contents of the cell \prodiden{string} can be defined using interpolation of values made
available in the context of the advice (for instance, join point contents) using
interpolation. For example, to add a row which evaluates the total of a
column (for instance, the join point is a column range) where the join point is named
\textit{myColumn}:

\medskip
\lstinline|below myColumn { =SUM(#{range.name}) }|
\medskip

In the above example, interpolation is used to introduce a value that is
available in the join point, but is not accessible from common spreadsheet
formulas. If the column range is \texttt{\small C1:C20}, then the formula would
be \texttt{\small =SUM(C1:C20)}.

Note that the order by which the advice are applied is important.
They are applied according to their precedence which
is defined by the order they are defined. In the case of \lstinline|before|
the ones defined first are the ones with more precedence. Thus,
the ones defined earlier are applied first. 
For the \lstinline|after| it is the other way around, that is,
although the ones declared first are the ones with more priority,
they are executed last. This is the common behavior of aspects 
for other programming languages.

Advice can be applied conditionally, that is, when a specified criterion is met.
This is specified with a boolean expression as defined by the following grammar:

\medskip
\begin{grammar}
<bExpr> ::= `!' <bExpr>
\alt <bExpr> (`&&' | `||') <bExpr>
\alt <expr> <bComp> <expr>

<expression> ::= <bExpr> |~<var> |~<string> |~<number> |~<cellRef>
|~<rangeRef>
\end{grammar}

\subsection{Aspect}

An aspect is composed of the definition of the pointcuts and the advice to apply
to them, as defined by the following grammar:

\medskip
\begin{grammar}
<aspect> ::= `aspect' <string>\\
             <join\_point>+ <advice>+\\
             `end'
\end{grammar}
\medskip

%
\section{A Weaver For Spreadsheets}
\label{sec:weaver}

In this section we present our architectural model for aspect \sds.  The aspects
are defined using the language presented in Section~\ref{sec:language}, and
together with the \sd, they are interpreted by the \textit{weaver}.  In this
context, the \sd is only complete when considered together with the aspects.

Some aspects can be handled statically, that is, without executing the \sd. For
instance, the example shown is Listing~\ref{lst:ectsmark} does not need to
execute the \sd to know what to do.

However, some aspects require the execution of the program, in our case, of the
\sd (formulas). This is the case illustrate in Listing~\ref{lst:borderline}
where it is necessary to \textit{compute} the student grade to decide if the
final grade is changed or not.  In that case we used the operator
\lstinline|cell.result|, which requires to evaluate the formula of the
underlying cell to obtain its result.  Thus, the weaver must be integrated with
a \sd execution engine.  Indeed we intend to build the weaver inside Excel
itself, so it can reuse its recalculation mechanism. 

In these cases, the \sd is kept untouched, but since now the spreadsheet is only
complete when considered with the aspects defined, the values it shows may
change. 

If the user wants to see the original \sd then, it is only necessary to
deactivate the aspects.

\begin{figure}[!htb]
\includegraphics[width=\columnwidth]{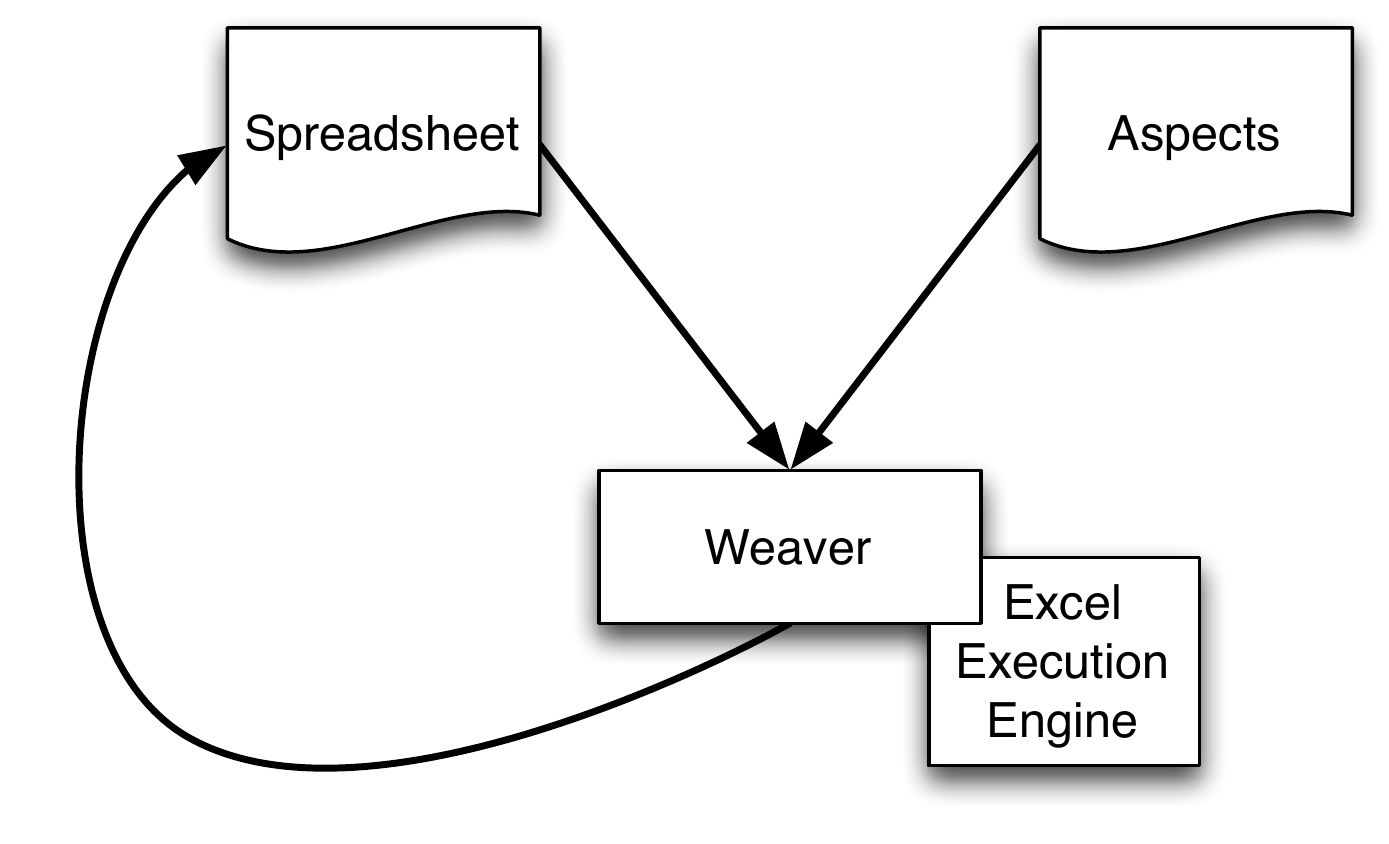}
\caption{A weaver for aspect-oriented spreadsheets.}
\label{fig:arc}
\end{figure}

Fig.~\ref{fig:arc} illustrates the integration of our weaver with a spreadsheet
system to create an aspect-oriented \sd system.

%
\section{Related Work}
\label{sec:related}

Aspect-oriented programming has been applied to several programming languages,
for instance, Java~\cite{kiczales2001overview}, C++~\cite{aspectc++2002}, or
\matlab~\cite{cardoso2010domain,aspectmatlab2010}, targeting specific
characteristics of each language in order to improve them. Some of these
applications contributed to shape the use of aspects, and introduced new
concepts to this paradigm.

Java was one of the first languages to be exposed to aspects through a language
dubbed AspectJ~\cite{kiczales2001overview}. This language is similar to Java so
it feels familiar to Java users. It provides a dynamic join point model, where
some advice are applied at compile time, but others are applied only during run
time when the complete information about the execution is available. The nature
of spreadsheets, where both data and computation are at the same level, imposes
a dynamic join point model in order to have access to run-time values.

AOP was also used to support development of embedded systems. The LARA
language~\cite{cardoso2012lara} was purposely designed with this goal, but has a
wider range of applications. Since it can target several languages, we inspired
ourselves on it as the basis for the specification of aspects for spreadsheets.

In the context of spreasheets, there are several works presenting techniques to
transform spreadsheets by using spreadsheet specific transformation languages
\cite{felienne2014bumblebee,badame2012refbook}, and querying languages
\cite{vlhcc-td13,vlhcc13}.  Moreover, such transformations can be refactorings
to remove spreadsheet
smells~\cite{felienne2012smells,Cunha:2012:TCS:2346340.2346357,
CunhaFMMS12,quatic14} and thus improve their usage and reduce possible error
entry points.

BumbleBee~\cite{felienne2014bumblebee} is a Microsoft Excel add-in mainly for
performing refactorings to remove smells, but can also perform other kinds of
transformations. It finds cells where it can apply a previously defined set of
transformations, lets the user select the transformation to apply, and then
applies the transformation to a selected range, to the entire worksheet, or to
the entire file. A limitation is that BumbleBee only supports intra-formula
transformations. Our language uses the BumbleBee's transformation language to
support cell value transformations.

RefBook~\cite{badame2012refbook} is another tool to perform refactorings to
remove spreadsheet smells. It implements a set of seven refactorings in a
Microsoft Excel add-in allowing users to perform refactorings when working with
their spreadsheets. Our approach can also be used to perform these refactorings.

More generic transformations for spreadsheets, introduced by end-users' needs,
can be performed using program synthesis. This technique added the ability to
transform strings of text~\cite{Gulwani2011}, or tables of
data~\cite{Harris2011} from user-supplied examples, providing a familiar way to
solve common tasks when dealing with spreadsheets.

Another kind of transformation, targeting spreadsheet testing, is
mutation~\cite{spreadsheetmutation}.  The goal is to perform mutations in the
spreadsheet, that is, small modifications, in order to analyze a test set for
the spreadsheet being tested. Using our aspect system, mutation is also
possible, by using the BumbleBee transformation language to specify cell
mutations.

In \cite{vlhcc10,ase14} we presented techniques to infer the business logic of
spreadsheet data. Such techniques restructure the spreadsheet data into
different (relational) tables. Such tables can be viewed as aspects of the
business logic/spreadsheet data. Thus, this approach may be used to infer
aspects in spreadsheets and to evolve a legacy spreadsheet into an
aspect-oriented one.

%
\section{Conclusion}
\label{sec:conclusion}

This paper proposes the use of aspect-oriented programming for spreadsheets. We
have designed an aspect language that considers spreadsheet peculiarities, and a
dynamic weaver that is embedded in the evaluation mechanism of a spreadsheet
system. 

Although AOP provides a powerful modular approach that is particularly suitable
to be used in software that is shared and being collaboratively developed, our
work opens some important questions that we intend to answer in future work by
conducting empirical studies, namely:

\begin{itemize}

\item Are end users able to understand the abstractions provided by AOP
  and to use it in practice?\\
  This is not only related to our proposed
  language as it is a more general question. Nevertheless, we need 
  to answer it so we can understand how to better make our AOP language
  available for user. This is specially important when dealing with end users.

\item Does AOP improves end-users' productivity?\\
  We have shown that some of the model-driven approaches we introduced
  in the past can do that. We will conduct similar studies to evaluate
  this new proposal.

\item Is the textual definition of the AOP language adequate or should
  we use a more spreadsheet-like one? \\
  When dealing with more advanced users, it is not always the case they
  prefer visual languages~\cite{2015arXiv150207948C}. However, for end users
  this is probably the case. We will extend our work with a visual
  language to allow end users to define aspects in a more friendly way.

\end{itemize}

%
%

\section*{Acknowledgment}
This work is financed by the FCT – Fundação para a Ciência e a Tecnologia 
(Portuguese Foundation for Science and Technology) within project UID/EEA/50014/2013.
This work has also been partially funded by FLAD/NSF
through a project grant (ref. 233/2014).
The last author is supported
  by CAPES through a \textit{Programa Professor Visitante do
      Exterior (PVE)} grant (ref. 15075133).

%
%

\IEEEtriggeratref{19}


\end{document}